\documentclass[aps,prb,reprint,amsmath,amssymb,showpacs,
groupedaddress]{revtex4-1}

\usepackage{graphicx}% Include figure files
\usepackage{dcolumn}% Align table columns on decimal point
\usepackage{bm}% bold math

\begin{document}

% Use the \preprint command to place your local institutional report
% number in the upper righthand corner of the title page in preprint mode.
% Multiple \preprint commands are allowed.
% Use the 'preprintnumbers' class option to override journal defaults
% to display numbers if necessary
%\preprint{}

%Title of paper
\title{Geometric resonances in the magnetoresistance of hexagonal lateral superlattices}

% repeat the \author .. \affiliation  etc. as needed
% \email, \thanks, \homepage, \altaffiliation all apply to the current
% author. Explanatory text should go in the []'s, actual e-mail
% address or url should go in the {}'s for \email and \homepage.
% Please use the appropriate macro foreach each type of information

% \affiliation command applies to all authors since the last
% \affiliation command. The \affiliation command should follow the
% other information
% \affiliation can be followed by \email, \homepage, \thanks as well.
\author{Yuto Kato}
%\email[]{Your e-mail address}
%\homepage[]{Your web page}
%\thanks{}
%\altaffiliation{}
\affiliation{The Institute for Solid State Physics, The University of Tokyo, 5-1-5 Kashiwanoha, Kashiwa, Chiba 277-8581, Japan}

\author{Akira Endo}
\email[]{Corresponding author: akrendo@issp.u-tokyo.ac.jp}
%\thanks{Corresponding author}
\affiliation{The Institute for Solid State Physics, The University of Tokyo, 5-1-5 Kashiwanoha, Kashiwa, Chiba 277-8581, Japan}

\author{Shingo Katsumoto}
\affiliation{The Institute for Solid State Physics, The University of Tokyo, 5-1-5 Kashiwanoha, Kashiwa, Chiba 277-8581, Japan}

\author{Yasuhiro Iye}
\affiliation{The Institute for Solid State Physics, The University of Tokyo, 5-1-5 Kashiwanoha, Kashiwa, Chiba 277-8581, Japan}

%Collaboration name if desired (requires use of superscriptaddress
%option in \documentclass). \noaffiliation is required (may also be
%used with the \author command).
%\collaboration can be followed by \email, \homepage, \thanks as well.
%\collaboration{}
%\noaffiliation

\date{\today}

\begin{abstract}
% insert abstract here
We have measured magnetoresistance of hexagonal lateral superlattices. We observe three types of oscillations engendered by periodic potential modulation having hexagonal-lattice symmetry: amplitude modulation of the Shubnikov-de Haas oscillations, commensurability oscillations, and the geometric resonances of open orbits generated by Bragg reflections. The latter two reveal the presence of two characteristic periodicities, $\sqrt{3} a / 2$ and $a / 2$, inherent in a hexagonal lattice with the lattice constant $a$. The formation of the hexagonal-superlattice minibands manifested by the observation of open orbits marks the first step toward realizing massless Dirac fermions in semiconductor 2DEGs. 
\end{abstract}

% insert suggested PACS numbers in braces on next line
% insert suggested keywords - APS authors don't need to do this
%\keywords{}

%\maketitle must follow title, authors, abstract, \pacs, and \keywords
\maketitle

% body of paper here - Use proper section commands
% References should be done using the \cite, \ref, and \label commands
\section{Introduction\label{Intro}}
% Put \label in argument of \section for cross-referencing

A hexagonal lateral superlattice (HLSL) --- a two-dimensional electron gas (2DEG) subjected to periodic potential modulation with hexagonal-lattice symmetry --- is of interest in two different contexts. First, it is envisaged as a route to artificially generate massless Dirac fermions (MDF) at the corners of the superlattice Brillouin zone. \cite{Park09N,Gibertini09,Simoni10,Nadvornik12,Gomes12} Second, it is expected to stabilize \cite{Haldane00} the fragile ``bubble phase'' (the hexagonal-lattice arrangement of two- or three-electron clusters) in the quantum Hall system, which has been theoretically predicted to be the ground state at the $1/4$ or $3/4$ fillings of the third or higher Landau levels. \cite{Koulakov96,Fogler96,Moessner96} Analogous stabilization is reported for the stripe phase at the half fillings, using one-dimensional (1D) lateral superlattices. \cite{Endo02f,Endo03lt} As an initial step toward pursuing these intriguing possibilities, we study, in the present work, low-field magnetoresistance of lateral superlattices with a weak (a few percent of the Fermi energy $E_\text{F}$) hexagonal-lattice potential modulation, fabricated from conventional GaAs/AlGaAs 2DEGs. Oscillations observed in the low-field magnetoresistance serve as a tool to characterize the HLSL samples we prepare.

Numerous studies have been devoted to lateral superlattices having 1D \cite{Weiss89,Beton90P,*[{See, e.g, }][{, and references therein.}] Endo05P} and two-dimensional (2D) square- \cite{Weiss90,Gerhardts91,Lorke91,Weiss92,Chowdhury00,Chowdhury04} or rectangular-lattice \cite{Chowdhury01,Geisler05} potential modulations. By comparison, magnetotransport of HLSLs remains relatively unexplored. An early experiment by Fang and Stiles \cite{Fang90} exhibited, for a weak modulation amplitude, commensurability oscillations (CO) similar to those observed in 1D lateral superlattices \cite{Weiss89} but corresponding to the periodicity half the lattice constant of the hexagonal lattice. Hexagonal lattices are not uncommon in antidots, \cite{Yamashiro91,Weiss94SS,Nihey95,Iye04JPSJ,Meckler05} which represent the strong limit of the modulation amplitudes. As demonstrated in Ref.\ \onlinecite{Fang90} (and also in Ref.\ \onlinecite{Lorke91} for square lattices), however, antidots and weakly modulated lateral superlattices display qualitatively distinct behavior. 

In the present paper, we report three variants of oscillatory phenomena in HLSLs qualitatively similar to those known in 1D lateral superlattices: CO
%(sometimes also called as ``geometric resonances'')
due to the commensurability between the cyclotron radius and the periodicity in the superlattice, \cite{Weiss89} amplitude modulation (AM) of the Shubnikov-de Haas oscillations (SdHO), \cite{OverendG98,Milton00,Edmonds01,Shi02,Endo08ModSdH} and an alternative type of oscillations, geometric resonances of open orbits (GROO). \cite{Endo05N,Endo06E,Endo08GR} Open orbits are composed of segments of cyclotron orbits repeatedly diffracted by Bragg reflections from the superlattice potential. Resonances take place when the width of the open orbits coincides with the periodicity. In both CO and GROO, oscillations are observed as the superposition of two components originating from the periodicities $\sqrt{3} a / 2$ and $a / 2$, respectively, with $a$ representing the lattice constant of the hexagonal lattice. The CO for the latter periodicity corresponds to that observed in Ref.\ \onlinecite{Fang90} mentioned above. We will discuss the amplitude of CO in connection with the amplitude of the potential modulation inferred from the AM of SdHO\@.

One of the major motivations in the exploration of superlattices is to artificially design and generate a band structure (miniband) that possesses the length and energy scale different from that in natural crystals, \cite{Esaki70} with the generation of MDF in HLSL being one example. In lateral superlattices, however, clear evidence of the formation of the miniband structure has not been observed until recently, \cite{Albrecht99,Deutschmann01} probably due to the technical difficulty in fabricating, with minimal disorder, superlattices with the period small enough (close to the Fermi wavelength) to embrace minibands. Note that both CO and AM of SdHO can be traced back to the oscillation with the magnetic field of the width of the Landau bands (Landau levels broadened by the modulation potential), and do not require minibands for their occurrence. By contrast, the observation of GROO evinces the formation of miniband structure, attesting to the high quality of our superlattice samples eligible to seek for artificial Dirac fermions.

The paper is organized as follows. After describing experimental details in Sec.\ \ref{Expdet}, experimentally obtained magnetoresistance traces exhibiting CO, AM of SdHO, and GROO are presented in Secs.\ \ref{ssCO}, \ref{ssSdH}, and \ref{ssGR}, respectively. Prospects and necessary improvements to be made to realize MDF and to detect the bubble phase are discussed in Secs.\ \ref{DiscMDF} and \ref{DiscBP}, respectively, followed by concluding remarks in Sec.\ \ref{Conc}.

\section{Experimental details\label{Expdet}}
\begin{figure}[tb]
\includegraphics[width=8.6cm]{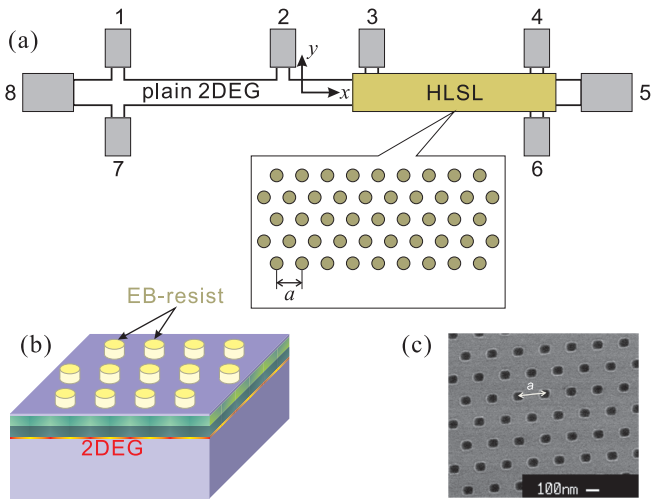}%
\caption{(Color online) Hexagonal lateral superlattice (HLSL) sample used in the present study. (a) Hall bar containing both the section with hexagonal-lattice potential modulation (HLSL) and the section without modulation (plain 2DEG) for reference. (b) Schematic drawing of the HLSL section. (c) Scanning electron micrograph of the hexagonal lattice of EB-resist (black dots) placed on the surface of the 2DEG wafer. \label{samplesch}}
\end{figure}
Schematics of the HLSL samples used in the present study are depicted in Fig.\ \ref{samplesch}. The samples were fabricated from a conventional GaAs/AlGaAs 2DEG wafer with the heterointerface residing at the depth $d = 60$ nm from the surface. As shown in Fig.\ \ref{samplesch}(a), the 2DEG wafer was patterned into Hall bar with the width 40 $\mu$m and containing two sets of voltage probes (with the inter-probe distance 320 $\mu$m) to measure the section with (HLSL) and without (plain 2DEG, for reference) the hexagonal-lattice  modulation. The potential modulation was introduced by placing a hexagonal lattice of high-resolution negative electron-beam (EB) resist (calixarene) \cite{Fujita96} on the surface (Fig.\ \ref{samplesch}(b)(c)) and making use of the strain-induced piezoelectric effect, \cite{Skuras97} as has been done to prepare 1D lateral superlattices. \cite{Endo00e,Endo02f,Endo03lt,Endo05HH,Endo05N,Endo06E,Endo08GR,Endo08ModSdH} 
Compared with more general methods to introduce potential modulations, e.g., by placing metallic gate grids or by shallow etching, the simpleness of our approach (only one-step process, the EB drawing, is needed to introduce the modulation), along with the high spatial resolution of the EB resist we employ, allows us to prepare highly ordered lateral superlattice samples with minimal damage to the 2DEG\@. 
In fact, the mobility $\mu = 88$ m$^2$V$^{-1}$s$^{-1}$ and the electron density $n_e = 3.9\times 10^{15}$ m$^{-2}$ of the 2DEG wafer remained virtually unchanged after the fabrication of the HLSL devices. We prepared HLSL samples with the lattice constant $a = 200$ nm and $100$ nm. Since the modulation amplitude for $a = 100$ nm was found to be extremely small, we mainly present the data taken from $a = 200$ nm HLSL in the followings. Note that the modulation strength rapidly decreases with decreasing $a$, roughly as $\exp (-a / d)$. \cite{Endo05HH} 
Resistivity measurements were carried out employing standard low-frequency ac lock-in technique at 4.2 K for CO and GROO, and at 15 mK, using a dilution refrigerator, for  SdHO\@.

\section{Experimental results}
\subsection{Commensurability oscillations\label{ssCO}}
\begin{figure}
\includegraphics[bbllx=10,bblly=68,bburx=380,bbury=795,width=8.6cm]{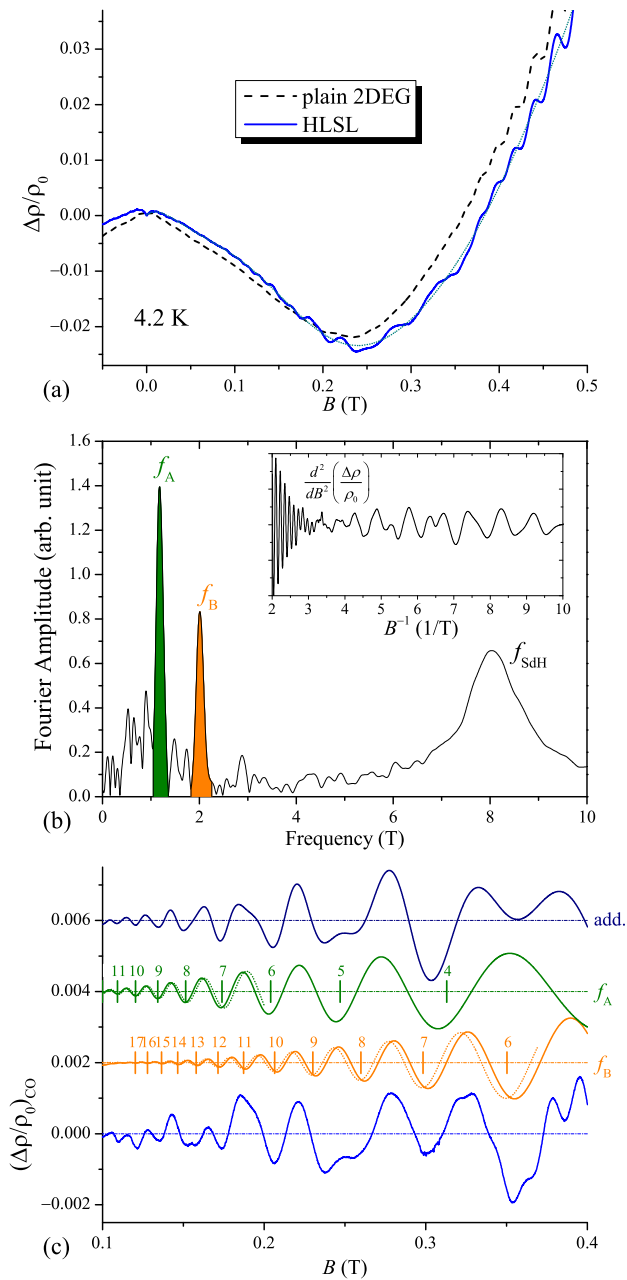}%
\caption{(Color online) (a) Magnetoresistance traces at 4.2 K for the HLSL with $a = 200$ nm (solid line) and the adjacent plain 2DEG (dashed line). Dotted curve represents slowly-varying background for the HLSL section, obtained by polynomial fitting. (b) Inset: Oscillatory part for the HLSL section obtained by taking the second derivative with respect to $B$, plotted against $B^{-1}$. Main: Fourier spectrum of the oscillatory part shown in the inset. (c) Oscillatory part of the magnetoresistance for the HLSL section obtained by subtracting the slowly-varying background shown in (a) (bottom), components having the frequency $f_\text{A}$ and $f_\text{B}$ extracted by the Fourier bandpass filter method (see text for details), and the addition of the $f_\text{A}$ and $f_\text{B}$ components (top). The latter traces are offset by $0.002$ for clarity. Vertical ticks for the $f_\text{A}$ and $f_\text{B}$ components indicate the positions of the $n$-th flat-band conditions given by Eq.\ (\ref{flatband}). Dotted curves are the fit to Eq.\ (\ref{CO}). \label{WeissAll}}
\end{figure}
\begin{figure}
\includegraphics[width=8.6cm]{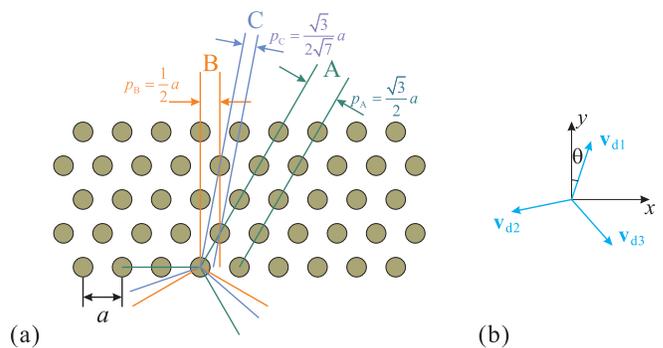}%
\caption{(Color online) (a) Representative lattice spacings in the hexagonal lattice. The periodicities $p_\text{A}$ and $p_\text{B}$ account for the components $f_\text{A}$ and $f_\text{B}$ in Fig.\ \ref{WeissAll}, respectively. Periodicities equivalent to those shown are found by rotating them by $\pm$120$^\circ$. (b) Drift velocity $v_\text{d1}$ along the direction $\theta$ from the $y$-axis and $v_\text{d2}$, $v_\text{d3}$ along the other two equivalent directions. $\theta = 30^\circ, 0^\circ$, and $\text{arccot}(3\sqrt{3}) \simeq 10.9^\circ$ for the periodicities $p_\text{A}$, $p_\text{B}$, and $p_\text{C}$, respectively. \label{PeriodsRot}}
\end{figure}
Magnetoresistance $\Delta \rho (B) / \rho_0$ of a HLSL with $a = 200$ nm is shown in Fig.\ \ref{WeissAll}(a). Here, $\rho_0$ is the resistivity $\rho(B)$ at $B=0$ and $\Delta \rho (B) \equiv \rho (B) - \rho_0$. Oscillatory behavior is apparent above $\sim$0.1 T (see also the bottom trace in Fig.\ \ref{WeissAll}(c)), which is absent in the trace for the plain 2DEG\@. Small oscillations seen above $\sim$0.35 T for both traces are the SdHO\@. To analyze the oscillations, we take the second derivative $(d^2/dB^2)(\Delta \rho (B) / \rho_0)$ numerically to extract the oscillatory part, plot it against $1/B$ (inset to Fig.\ \ref{WeissAll}(b)), and then perform Fourier transform. The Fourier spectrum thus obtained is presented in the main panel of Fig.\ \ref{WeissAll}(b). Three eminent peaks, $f_\text{A}$, $f_\text{B}$, and $f_\text{SdH}$, are seen: $f_\text{SdH} = n_e h / (2e)$ represents the SdHO, while $f_\text{A}$ and $f_\text{B}$ coincide with the frequency $2 \hbar k_\text{F} / (e p)$, with $k_\text{F} = \sqrt{2 \pi n_e}$ the Fermi wave number, of the CO corresponding to the periodicities $p = \sqrt{3} a / 2 \equiv p_\text{A}$ and $a / 2 \equiv p_\text{B}$, respectively. As depicted in Fig.\ \ref{PeriodsRot}(a), both $p_\text{A}$ and $p_\text{B}$ are the representative spacings between the lattice points contained in the hexagonal lattice. More generally, lattice spacings in the hexagonal lattice are given by $p_{(h,k)} = 2 \pi / |\bm{g}_{(h,k)}|$, with $p_\text{A}$, $p_\text{B}$, and $p_\text{C}$ corresponding to $(h,k) = (1,0)$, $(1,1)$, and $(2,1)$, respectively. Here, $\bm{g}_{(h,k)} = h \bm{a_1}^* + k \bm{a_2}^*$ represents a reciprocal lattice vector with $\bm{a_1}^* = (2 \pi / a)(1,-1/\sqrt{3})$ and $\bm{a_2}^* = (2 \pi /a )(1,1/\sqrt{3})$ the primitive reciprocal lattice vectors. A potential modulation having the hexagonal-lattice symmetry can generally be written as,
\begin{equation}
V(\bm{r}) = \sum\limits_{h,k} {V_{p_{(h,k)}} \cos (\bm{g}_{(h,k)}  \cdot \bm{r})},
\label{PotMod}
\end{equation}
with $\bm{r} = (x,y)$. We defined the direction of the current in the measurement of the resistance as the $x$ axis (see Fig.\ \ref{samplesch}(a)). The summation is taken over sets of integers $(h,k)$ that yield independent periodicities in the 2D plane. \footnote{Double counting can conveniently be avoided, e.g., by restricting to $h \geq 0$, with $k \gtrless 0$ for $h > 0$, and $k >0 $ for $h = 0$.}  
Hexagonal modulation potential often given in the literatures (e.g., Eq.\ (3) in Ref.\ \onlinecite{Fang90} and Eq.\ (1) in Ref.\ \onlinecite{Nadvornik12}, see also Eq.\ (\ref{VpA}) below) corresponds to Eq.\ (\ref{PotMod}) retaining only the components with the largest spacings $p_\text{A}$, namely, $V_{p_{(h,k)}}$ with $(h,k) = (1,0), (1,-1), (0,-1)$.
The presence of the two peaks $f_\text{A}$ and $f_\text{B}$ in the Fourier spectrum reveals that the observed CO is the superposition of two components corresponding to the periodicities $p_\text{A}$ and $p_\text{B}$, and that the components corresponding the periodicity $p_\text{B}$, $V_{p_{(h,k)}}$ with $(h,k) = (1,1), (2,-1), (1,-2)$, are also present in the modulation potential Eq.\ (\ref{PotMod}).

To gain more quantitative information from the CO, we separate out individual components following the prescription described in Ref.\ \onlinecite{Endo08FCO}: first, we perform Fourier band-pass filter on $(d^2/dB^2)(\Delta \rho (B) / \rho_0)$ vs. $B^{-1}$ plotted in the inset of Fig.\ \ref{WeissAll}, employing the window that encompasses the peak $f_\text{A}$ or $f_\text{B}$ (shaded areas in Fig.\ \ref{WeissAll}(b)); $(d^2/dB^2)(\Delta \rho (B) / \rho_0)$ corresponding to a single component thus extracted are replotted against $B$ and then numerically integrated by $B$ twice. The traces of single component $(\Delta \rho (B) / \rho_0)_\text{CO}$ for $f_\text{A}$ and $f_\text{B}$ restored by this procedure are plotted in Fig.\ \ref{WeissAll}(c). As demonstrated in the figure, addition of the $f_\text{A}$ and $f_\text{B}$ components reproduces well the oscillatory part of $\Delta \rho (B) / \rho_0$ directly obtained by subtracting the slowly varying background found by polynomial fitting to the data (plotted with the dotted line in Fig.\ \ref{WeissAll}(a)), apart from the SdHO above $\sim$0.35 T\@. In Fig.\ \ref{WeissAll}(c), positions of the flat-band conditions, where the drift velocity $v_\text{d}$ vanishes,
\begin{equation}
\frac{2 R_\text{c}}{p} = n-\frac{1}{4} \hspace{10mm} (n = 1,2,3,...),
\label{flatband}
\end{equation}
are indicated by vertical ticks. Here, $R_\text{c} = \hbar k_\text{F} / (eB)$ is the cyclotron radius. It can be seen for both components that the minima take place at the flat-band conditions, as is the case with 1D lateral superlattices. \cite{Weiss89}$^,$\footnote{Slight deviation from the exact positions given by Eq.\ (\ref{flatband}) is often seen also in 1D lateral superlattices.} To be more precise, the diffusion (band) contribution resulting from the drift velocity $v_\text{d}$ takes minima, while the collisional (hopping) contribution due to the modulation of the density of states (DOS) takes maxima, at Eq.\ (\ref{flatband}) in 1D lateral superlattices, \cite{Zhang90,Peeters92} with the former being by far the dominant contribution in the measurement of the resistivity along the principal axis of the modulation.

We further make an attempt to fit the extracted CO curves to a formula representing the diffusion contribution for 1D lateral superlattices $V(x) = V_p \cos (2 \pi x / p)$, in which the damping of the oscillations by scatterings is taken into account: \cite{Mirlin98,Endo00e}
\begin{equation}
\left( \frac{\Delta \rho}{\rho_0} \right)_\text{CO} \! \! = \gamma A\left( \frac{T}{T_p} \right) A\left( \frac{\pi}{\mu_\text{w} B} \right) \frac{2 \pi}{p} V_p^2 B \sin \left(2 \pi \frac{2 R_\text{c}}{p} \right),
\label{CO}
\end{equation}
where
\begin{equation}
\gamma=\frac{1}{2(2\pi)^{3/2}}\left(\frac{h}{e}\right)^{-1}\left(\frac{e\hbar}{2m^*}\right)^{-2}\frac{\mu^2}{n_e ^{3/2}},
\label{gamma}
\end{equation}
and $T_p \equiv p k_\text{F} \hbar \omega_\text{c} / (4 \pi^2 k_\text{B})$, with $m^*$ the effective mass, $k_\text{B}$ the Boltzmann constant, $\omega_\text{c} = eB / m^*$ the cyclotron angular frequency, and $A(X) \equiv X / \sinh X$. We use the modulation amplitude $V_p$ and the effective mobility $\mu_\text{w}$ as fitting parameters. As shown by dotted curves in Fig.\ \ref{WeissAll}(c), fairly good fitting is achieved, albeit within a rather limited magnetic-field range: $B \alt 0.20$ T for $f_\text{A}$ and $B \alt 0.37$ T for $f_\text{B}$. \footnote{Above this $B$ range, the oscillation amplitude grows with $B$ slower than predicted by Eq.\ (\ref{CO}).} The values of $V_p$ and $\mu_\text{w}$ obtained by the fittings are $V_p = 0.025$ and $0.016$ meV, and  $\mu_\text{w} = 9.1$ and $8.2$ m$^2$V$^{-1}$s$^{-1}$ for $f_\text{A}$ and $f_\text{B}$ ($p = p_\text{A} = 173$ nm and $p_\text{B} = 100$ nm), respectively. It has been shown for 1D lateral superlattices \cite{Endo00e} that $\mu_\text{w}$ is close to the single particle (or quantum) mobility $\mu_\text{s}$ that describes the damping $\propto \exp(-\pi /\mu_\text{s} B)$ of the SdHO\@. \cite{Coleridge91} This is found to be roughly the case also for our HLSL; we obtain $\mu_\text{s} = 6.2$ m$^2$V$^{-1}$s$^{-1}$ from the similar Fourier analysis of SdHO for the data shown in Fig.\ \ref{WeissAll}(a) (and also for the data taken at $T \sim 15$ mK, see Fig.\ \ref{calccmpSdH}; dependence of $\mu_\text{s}$ on the temperature was not observed in this temperature range). The values of $V_p$, on the other hand, appear to be too small. Much larger values were found for 1D lateral superlattices having similar modulation periods and fabricated from the same 2DEG wafer and therefore expected to have similar modulation amplitudes: 0.18 and 0.06 meV for the periods $a = 200$ and 90 nm, respectively. \cite{Koike12,Kajioka11MT} The $V_p$ obtained here for the HLSL cannot be literally taken to represent the modulation amplitude in Eq.\ (\ref{PotMod}) correctly for the following reasons. 

First, it is necessary to note that the observed CO is an addition of contributions from three equivalent modulations with the same spacing present in the hexagonal lattice, rotated by 120$^\circ$ from each other. The drift velocity $v_\text{d}(\propto V_p)$ responsible for the diffusion contribution is pointed perpendicular to the direction of the modulation axis. As depicted in Fig.\ \ref{PeriodsRot}(b), the drift velocity is directed toward the angle $\theta$ ($\mathbf{v}_\text{d1}$) and $\theta \pm 120^\circ$ ($\mathbf{v}_\text{d2}$, $\mathbf{v}_\text{d3}$) from the $y$ axis,
%where we defined the direction we measure the resistivity $\rho_{xx}$ as the $x$ axis,
with $\theta = 30^\circ$ and $0^\circ$ for the modulations A and B, respectively. Since $\rho_{xx} \simeq \sigma_{yy}/\sigma_{yx}^2$ (with $\sigma_{\alpha \beta}$ representing components of the conductivity tensor), CO is proportional to the $y$ component of the drift velocity squared, $\Delta \rho_\text{CO} \propto |\mathbf{v}_\text{d1}|^2\cos^2\theta+|\mathbf{v}_\text{d2}|^2\cos^2(\theta+120^\circ)+|\mathbf{v}_\text{d3}|^2\cos^2(\theta-120^\circ)$. This equals $(3/2)v_\text{d}^2$ regardless of the angle $\theta$ if we assume $|\mathbf{v}_\text{d1}|=|\mathbf{v}_\text{d2}|=|\mathbf{v}_\text{d3}| \equiv v_\text{d}$, \footnote{This is equivalent to assuming that the amplitude $V_p$ is the same for the three directions, e.g., $V_{(1,0)} = V_{(1,-1)} = V_{(0,-1)} = V_{p_\text{A}}$ and $V_{(1,1)} = V_{(2,-1)} = V_{(0,-2)} = V_{p_\text{B}}$. This is not obvious in our sample, since potential modulation is introduced by the piezoelectric effect, which depends on the crystallographic directions. (In our Hall bars, $x$ axis is set parallel to one of the $<110>$ directions, the directions with the most prominent piezoelectric effect.) The effect of the possible anisotropy in the modulation amplitude is too complicated to be discussed with the data available in the present study.} leading to the correction of $V_p$ to the factor of $\sqrt{2/3}$ smaller values.

More importantly, it has been shown that the amplitudes of CO for 2D lateral superlattices are usually much smaller than those for 1D lateral superlattices having a similar modulation amplitude. \cite{Gerhardts91} This was initially attributed to the splitting of the Landau levels into sublevels (Hofstadter spectrum \cite{Hofstadter76,Claro79}), which suppresses the diffusion contribution. \cite{Gerhardts91,Weiss92} Later, an alternative explanation was presented by Grant \textit{et al}. \cite{Grant00} based on the calculation of semiclassical trajectories of the electrons showing that the drifting motion introduced by the modulation in the $x$ direction is suppressed by the modulation in the $y$ direction. They showed that the diffusion contribution survives, with the amplitude reduced compared to 1D modulation, only when the modulation is asymmetric between $x$ and $y$ directions. For symmetric modulation, the diffusion contribution vanishes, leaving only the collisional contribution having the oscillation phase opposite to that of the diffusion contribution. The effect of the symmetry between $x$ and $y$ directions was experimentally confirmed. \cite{Chowdhury00} Although these are for a 2D square lattice, qualitatively similar mechanism is expected to be operative also in the hexagonal lattice, in which modulations with differing orientations coexist. Therefore, the value of $V_p$ obtained by fitting Eq.\ (\ref{CO}) to the CO trace is expected to be smaller than the modulation amplitude also in the hexagonal lattice. This is confirmed by comparing $V_p$ obtained here to the modulation amplitude inferred from the AM of the SdHO, as will be shown in the subsequent subsection \ref{ssSdH}.

\subsection{Amplitude modulation of Shubnikov-de Haas oscillations\label{ssSdH}}
\begin{figure}
\includegraphics[bbllx=10,bblly=145,bburx=380,bbury=775,width=8.6cm]{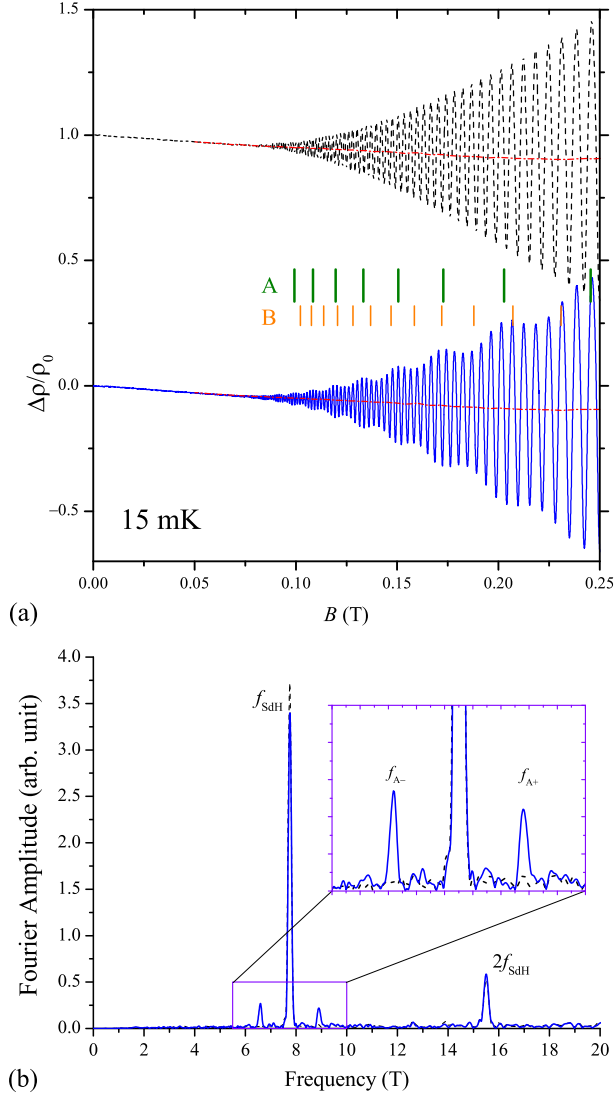}%
\caption{(Color online) Magnetoresistance traces at 15 mK for the HLSL with $a = 200$ nm  (solid line) and the adjacent plain 2DEG (dashed line, offset by 1.0 for clarity), showing rapid SdHO\@. Slowly-varying background obtained by averaging the upper and lower envelop curves of the $\Delta \rho / \rho_0$ for the plain 2DEG is shown by dot-dashed line. (The same background is shown for both traces). Vertical ticks indicate the flat-band conditions, Eq.\ (\ref{flatband}), for periodicities $p_\text{A}$ and $p_\text{B}$. (b) Fourier spectra taken of $(d^2/dB^2) (\Delta \rho / \rho_0)$ vs. $B^{-1}$ for the HLSL (solid line) and the plain 2DEG (dashed line) sections. \label{SdHAll}}
\end{figure}
\begin{figure}
\includegraphics[bbllx=10,bblly=30,bburx=720,bbury=530,width=8.6cm]{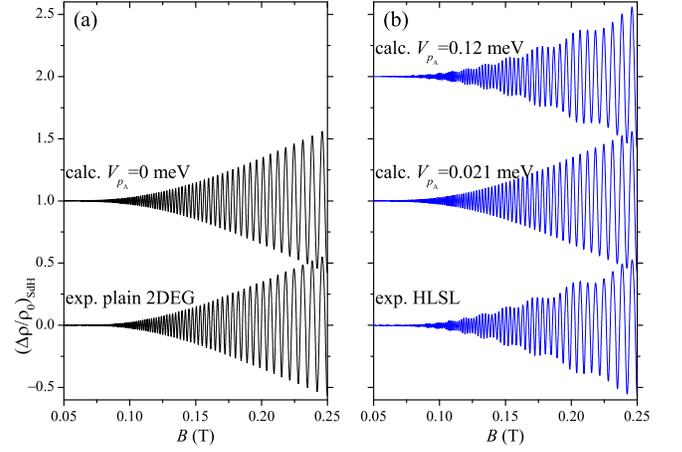}%
\caption{(Color online) Oscillatory parts of the magnetoresistance shown in Fig.\ \ref{SdHAll} obtained by subtracting the slowly-varying background (bottom) and those calculated by Eqs.\ (\ref{SdHcol}) and (\ref{VBHx}) with $n_e = 3.9 \times 10^{15}$ m$^{-2}$, $\mu_\text{s} = 6.2$ m$^2$V$^{-1}$s$^{-1}$, $T = 15$ mK, $C=2.2$, $p_\text{A} = 173$ nm, and the indicated values of $V_{p_\text{A}}$ for the plain 2DEG (a) and the HLSL (b) sections. Calculated traces are offset by 1.0 for clarity. \label{calccmpSdH}}
\end{figure}
In Fig.\ \ref{SdHAll}(a), we plot $\Delta \rho / \rho_0$ taken at $T = 15$ mK\@. It can readily be seen that the SdHO for the HLSL exhibits modulation in the oscillation amplitude, with the amplitude maxima at the flat-band conditions Eq.\ (\ref{flatband}) for the periodicity $p_\text{A} = \sqrt{3} a /2$. The AM is also evident in the Fourier spectrum shown in Fig.\ \ref{SdHAll}(b) taken of $(d^2/dB^2) (\Delta \rho / \rho_0)$ vs. $B^{-1}$, which exhibits, in addition to the peak representing the principal frequency $f_\text{SdH}$ of the SdHO and its harmonics, side peaks marked with $f_{\text{A}+}$ and $f_{\text{A}-}$, with the distance $|f_{\text{A}\pm}-f_\text{SdH}|$ being equal to the frequency of the CO, $2 \hbar k_\text{F} / (e p_\text{A})$, corresponding to the periodicity $p_\text{A}$; the distance coincides with $f_\text{A}$ in Fig.\ \ref{WeissAll}(b) after the correction of the small difference in the electron density $n_e$ between different cooling downs. The AM attributable to the periodicity $p_\text{B}$ was not clearly observed, probably owing to the weakness of the modulation. Note that, as mentioned earlier, the amplitude of the potential modulation rapidly decreases with decreasing periodicity. The absence of the $p_\text{B}$ component indicates that the AM of SdHO is more heavily weighted to larger amplitude components of the potential modulation compared to the CO\@.

The AM of SdHO is known to also originate from the two mechanisms, the diffusion contribution and the collisional contribution, with the amplitude minima (maxima) taking place at the flat-band conditions for the former (latter) mechanism. \cite{Zhang90,Peeters92,Endo08ModSdH} For 1D lateral superlattices, it has been shown that the collisional contribution dominates at low magnetic fields ($\alt 0.25$ T). \cite{Endo08ModSdH} This appears to be also the case in our HLSL, as can be seen in Fig.\ \ref{SdHAll}(a) exhibiting amplitude maxima at the flat band conditions. The diffusion contribution in the SdHO in HLSL, if any, is expected to be much smaller than in 1D lateral superlattices, by analogy with the case for the CO discussed above.

The collisional contribution to SdHO in 1D lateral superlattices $V(x) = V_p \cos (2 \pi x / p)$ is described well by, \cite{Endo08ModSdH}
\begin{eqnarray}
\label{SdHcol}
\left( \frac{\Delta \rho}{\rho_0} \right)_\text{SdH} ^\text{col} \! \! \! \! = -2C A\left( \frac{T}{T_\text{c}} \right) \hspace*{40mm}  \\
\times \exp \left( -\frac{\pi}{\mu_\text{s} B} \right) J_0 \left( \frac{2 \pi V_B}{\hbar \omega_\text{c}} \right)  \cos \left( \frac{2 \pi E_\text{F}}{\hbar \omega_\text{c}} \right), \nonumber
\end{eqnarray}
with 
\begin{equation}
V_B = V_p \frac{1}{\pi} \sqrt{\frac{p}{R_\text{c}}} \cos \left( \frac{2 \pi}{p} R_\text{c}-\frac{\pi}{4} \right)
\label{VB}
\end{equation}
representing the Landau bandwidth, $T_\text{c} \equiv \hbar \omega_\text{c} / (2 \pi^2 k_\text{B})$, $C$ a constant $\sim$2, \footnote{$C=2$ for ideally uniform 2DEGs but deviates from 2 in 2DEGs with small (a few percent) inhomogeneity in the electron density. See Ref.\ \onlinecite{Coleridge91}} and $J_0(x)$ the Bessel function of order zero. According to Wang \textit{et al.},\cite{Wang04} bandwidth of the $N$-th Landau level for a 2DEG subjected to the hexagonal potential modulation
\footnote{Equation (\ref{VpA}) corresponds to Eq.\ (\ref{PotMod}) with $V_{p_{(1,0)}} = V_{p_{(1,-1)}} = V_{p_{(0,-1)}} = V_{p_\text{A}}$ and $V_{p_{(h,k)}} = 0$ for all the other $(h,k)$.}
\begin{equation}
V(\bm{r}) = V_{p_\text{A}} 
\left[ \cos (\bm{a_1}^* \cdot \bm{r}) + \cos (\bm{a_2}^* \cdot \bm{r}) + \cos (\bm{a_3}^*\cdot \bm{r})\right]
\label{VpA}
\end{equation}
with $\bm{a_3}^* \equiv \bm{a_1}^* - \bm{a_2}^*$ is given by
$\frac{9}{4} V_{p_\text{A}} e ^ {-u/2} L_N (u)$, where $u = (2 \pi / p_\text{A})^2 l^2 / 2$ with $l = \sqrt{\hbar / (eB)}$ the magnetic length, and $L_N(u)$ represents the Laguerre polynomial.
The Landau bandwidth at the Fermi energy at a low magnetic field ($N \gg 1$) is thus approximated well by simply $9/4$ times Eq.\ (\ref{VB}):
\begin{equation}
V_B = \frac{9}{4} V_{p_\text{A}} \frac{1}{\pi} \sqrt{\frac{p_\text{A}}{R_\text{c}}} \cos \left( \frac{2 \pi}{p_\text{A}} R_\text{c}-\frac{\pi}{4} \right).
\label{VBHx}
\end{equation}
We therefore make an attempt to analyze AM of SdHO shown in Fig.\ \ref{SdHAll}(a) using Eqs.\ (\ref{SdHcol}) and (\ref{VBHx}).
In Fig.\ \ref{calccmpSdH}, we compare traces calculated by Eqs.\ (\ref{SdHcol}), (\ref{VBHx}) with SdHO extracted from the experimental $\Delta \rho / \rho_0$ shown in Fig.\ \ref{SdHAll}(a) by subtracting the slowly-varying background. As can be seen in Fig.\ \ref{calccmpSdH}(a), the trace calculated with $V_{p_\text{A}} = 0$ reproduces the experimental trace for the plain 2DEG section quite well. Figure \ref{calccmpSdH}(b) shows that AM is barely visible if we use the value $V_{p_\text{A}} = 0.021$ meV obtained by fitting the CO trace to Eq.\ (\ref{CO}) and applying the correction for the factor $\sqrt{2/3}$ to account for three different directions of the drift velocity outlined in Sec.\ \ref{ssCO}. To reproduce experimental AM, a much larger value $V_{p_\text{A}} = 0.12$ meV is required. Note that Eq.\ (\ref{VBHx}) already includes the contribution from the three equivalent orientations (see Eq.\ (\ref{VpA})). Since the collisional contribution is the effect of modulated DOS that alters the scattering rate of electrons, it essentially does not depend on the direction of the modulation, \cite{Zhang90,Peeters92} and therefore interference between different directions as in the diffusion contribution is considered to be absent. We thus expect the value of $V_p$ derived from the analysis of the SdHO presented here to represent better the amplitude of the modulation. The value is also consistent with the modulation amplitude of 1D lateral superlattices with similar modulation periods mentioned earlier.

\subsection{Geometric resonances of open orbits\label{ssGR}}
\begin{figure}
\includegraphics[bbllx=10,bblly=270,bburx=380,bbury=790,width=8.6cm]{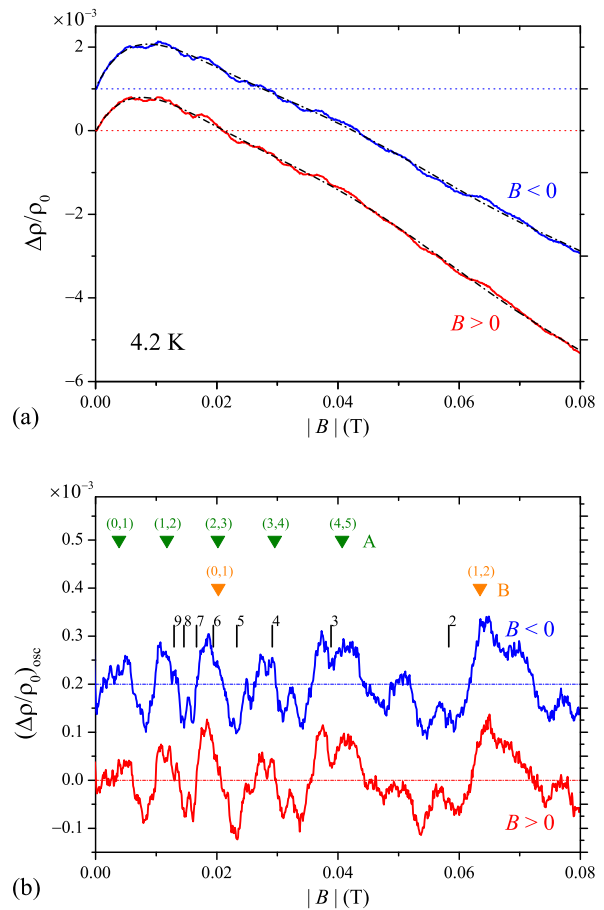}%
\caption{(Color online) (a) Low-field magnetoresistance traces of the HLSL section ($a = 200$ nm) at 4.2 K for $B>0$ and $B<0$ (offset by 1.0$\times$10$^{-3}$), plotted against the absolute value $|B|$. Slowly-varying backgrounds acquired by polynomial fitting are shown by dot-dashed lines. (b) Oscillatory parts for $B>0$ and $B<0$ (offset by 0.2$\times$10$^{-3}$), obtained by subtracting the slowly-varying background. Downward triangles mark the positions for the transverse resonance of the orbit X$(j,k)$ (X $=$ A or B), Eq.\ (\ref{Bt}). Vertical ticks indicate the positions for the $n$-th longitudinal resonance, Eq.\ (\ref{BlAB}). \label{GRAll}}
\end{figure}
\begin{figure}
\includegraphics[width=8.6cm]{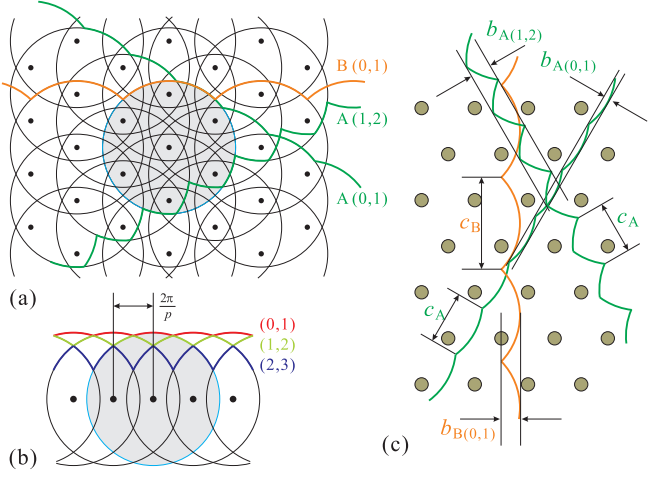}%
\caption{(Color online) Open orbits in the hexagonal lateral superlattice. (a) Open orbits in the reciprocal space. Open orbits generated by the periodicities $p_\text{A}$ , A$(0,1)$, A$(1,2)$, and by $p_\text{B}$, B$(0,1)$, are shown by thick lines. Black dots and circles represent the reciprocal lattice points and the Fermi circles, respectively. (b) Illustration in the reciprocal space of the index $(j,k)$ of the open orbits. (c) Open orbits in the real space. Width $b$ and periodicity along the length $c$ are shown. \label{GRIllust}}
\end{figure}
\begin{figure}
\includegraphics[width=8.6cm]{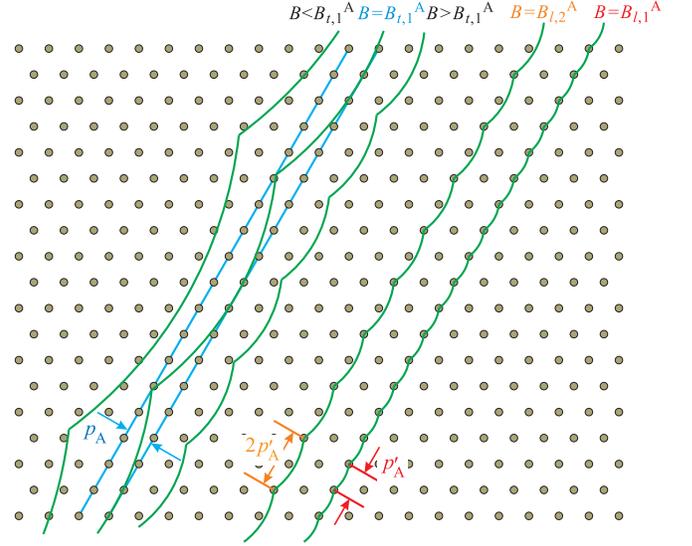}%
\caption{(Color online) Resonance conditions for the open orbit A$(0,1)$ generated by the periodicity $p_\text{A}$. Resonance takes place when the width $b_\text{A}$ of the open orbits coincides with the periodicity $p_\text{A}$ responsible for the generation of the orbit (\textit{transverse} resonance, $B=B_t$) or when the period $c_\text{A}$ along the length of the orbit matches the periodicity of the modulation $p^\prime_\mathrm{A}$ in the direction of the propagation (\textit{longitudinal} resonance, $B=B_l$). \label{GRrealRes}}
\end{figure}
We find yet another type of oscillations in the magnetic-field range below the regime where CO is observed. The small amplitude oscillations become clearly visible after the subtraction of slowly-varying background, as illustrated in Fig.\ \ref{GRAll}. Figure \ref{GRAll}(b) reveals that the oscillations observed at $B > 0$ are reproduced, including minute structures, in the trace taken at $B < 0$, ruling out the possibility that they simply result from  external noises. Since the measurement was performed on a large ($40 \times 320$ $\mu$m$^2$) Hall bar at a relatively high temperature, $T = 4.2$ K, the oscillations are unlikely to be related to the well-known universal conductance fluctuations (UCF). \cite{LeeStone85,Thornton87} The patterns of the oscillations do not change between separate cooling downs (see Fig.\ \ref{GR100} below), which is also at variance with UCF\@. Similar small amplitude oscillations have been reported in 1D lateral superlattices, and interpreted as the geometric resonances of open orbits (GROO). \cite{Endo05N,Endo06E,Endo08GR} The oscillations observed here in HLSL can also be interpreted with the same mechanism.

Open orbits are generated by Bragg reflections due to the superlattice potential. Figure \ref{GRIllust}(a) shows Fermi circles in the reciprocal space in the extended zone scheme, with the open orbits depicted as chains of Fermi circle segments (thick curves). As illustrated in Figs.\ \ref{GRIllust}(a) and (b), we denote the open orbit composed of arcs from $j$-th and $k$-th nearest-neighbor Fermi circles as X$(j,k)$, with X $=$ A, B for the open orbits engendered by the periodicities $p_\text{A}$ and $p_\text{B}$, respectively. The open orbit X$(j,k)$ is generated by the Bragg reflections from $j$-th and $k$-th harmonics of the modulation potential with the periodicity $p_\text{X}$. Open orbits in the real space are obtained by rotating those in the reciprocal space by 90$^\circ$ and multiplying by the factor $\hbar / (eB)$, as portrayed in Fig.\ \ref{GRIllust}(c). The width $b_{\text{X}(j,k)}$ of the orbit X$(j,k)$, therefore, decreases inversely proportional to $B$. The \textit{transverse} resonance takes place, leading to the maxima in the conductivity (hence in the resistivity as well), at magnetic fields, \cite{Endo05N}
\begin{equation}
{B_{t,n}}^\text{X} (j,k) = \frac{\hbar k_\text{F}}{n p_\text{X} e} \left[ \sqrt{1-\left( \frac{j \pi}{p_\text{X} k_\text{F}} \right)^2 } - \sqrt{1-\left( \frac{k \pi}{p_\text{X} k_\text{F}} \right)^2 } \right],
\label{Bt}
\end{equation}
when the width $b_{\text{X}(j,k)}$ coincides with $n$ times the periodicity $p_\text{X}$, as depicted in Fig.\ \ref{GRrealRes} (left) for $n = 1$, taking the open orbit A$(0,1)$ generated by the periodicity $p_\text{A}$ as an example. The positions given by Eq.\ (\ref{Bt}) with $n = 1$ are indicated by downward triangles in Fig.\ \ref{GRAll}(b), showing that the major features are explicable with the transverse resonances with lower indices $(j,k)$ for the periodicity either $p_\text{A}$ or $p_\text{B}$.

Since the open orbits possess the periodicity $c_\text{X} = h / (e B p_\text{X})$ (regardless of the indices $(j,k)$) in the direction of the propagation as illustrated in Fig.\ \ref{GRIllust}(c), and the periodicity of the potential modulation, $p^\prime_\text{X}$, is present also in that direction ($p^\prime_\text{A} = a$ and $p^\prime_\text{B} = \sqrt{3} a$), an alternative type of the resonances, \textit{longitudinal} resonances, can be considered when $c_\text{X}$ equals $n \cdot p^\prime_\text{X}$, namely at  
\begin{equation}
{B_{l,n}}^\text{X} = \frac{h}{n p_\text{X} p^\prime_\text{X} e},
\label{Bl}
\end{equation}
as exemplified in Fig.\ \ref{GRrealRes} (right). Note that Eq.\ (\ref{Bl}) reduces to
\begin{equation}
{B_{l,n}} = \frac{2}{\sqrt{3}} \frac{h}{n a^2 e}
\label{BlAB}
\end{equation}
for both the periodicities $p_\text{A}$ and $p_\text{B}$ in HLSL\@.
A similar resonance has been reported in a 1D lateral superlattice slightly modified to have the modulation also along the direction perpendicular to the principal axis (strictly speaking, therefore, it is a 2D rectangular lateral superlattice), showing a resistivity maximum at the condition corresponding to Eq.\ (\ref{Bl}). \cite{Endo05N} The positions of the resonances described by Eq.\ (\ref{BlAB}) are marked by vertical ticks in Fig.\ \ref{GRAll}(b), suggesting that some of the minor structures in the magnetoresistance are actually originating from the longitudinal resonances. Remnant unidentified minor structures could possibly be resulting from higher-order terms ($n \geq 2$) in Eq.\ (\ref{Bt}) for periodicities $p_\text{A}$ and $p_\text{B}$, or from the transverse resonances for still smaller lattice spacings embedded in the hexagonal lattice exemplified by $p_\text{C}$ in Fig.\ \ref{PeriodsRot}(a). (Positions for longitudinal resonances are still described by Eq.\ (\ref{BlAB}) for such periodicities.) However, these resonances are so densely distributed in the magnetic-field range shown in Fig.\ \ref{GRAll} that it is rather difficult to unambiguously identify the small features in Fig.\ \ref{GRAll}(b) with these resonances within the resolution of the present experiment.

\begin{figure}
\includegraphics[bbllx=40,bblly=25,bburx=750,bbury=510,width=8.6cm]{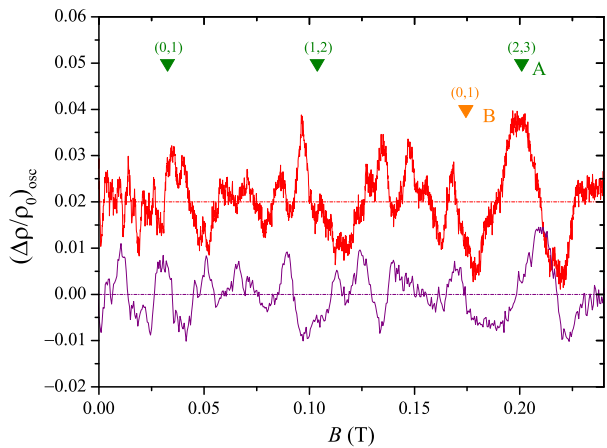}%
\caption{(Color online) Oscillatory part of the low-field magnetoresistance for the HLSL with $a = 100$ nm obtained by subtracting slowly-varying polynomial background.
Downward triangles indicate the positions for transverse resonances given by Eq.\ (\ref{Bt}). The lower trace is taken with $dB/dt = 1$ mT/s. The upper trace, offset by 0.02 for clarity, is taken on a different cooling down roughly one month after the bottom trace was acquired, with 5 times slower sweep rate $dB/dt = 0.2$ mT/s thus exhibiting higher resolution. \label{GR100}}
\end{figure}
Finally, we plot in Fig.\ \ref{GR100} the oscillatory part of low-field magnetoresistance for a HLSL with $a = 100$ nm analogous to that shown in Fig.\ \ref{GRAll}(b) for $a = 200$ nm. We found that both CO and AM of SdHO are extremely small for $a = 100$ nm, hindering us from drawing out reliable information on the potential modulation through the analyses similar to those done for $a = 200$ nm in the preceding subsections \ref{ssCO} and \ref{ssSdH}. This is attributed to the weakness of the potential modulation, owing to the smallness of the period $a$. Compared with these oscillations, GROO can be identified clearer as demonstrated in Fig.\ \ref{GR100}. This is consistent with the higher sensitivity of GROO, compared with CO, to a small periodicity $p$, mainly resulting from a higher characteristic temperature that governs the decrease of the oscillation amplitude with increasing temperature or decreasing magnetic field. \cite{Endo06E} In Fig.\ \ref{GR100}, two traces taken on different cooling downs are shown. Major peaks clearly recognizable in both traces take place at the positions, indicated by the downward triangles, of the transverse resonances Eq.\ (\ref{Bt}) with $n = 1$ having lower indices $(j,k)$ for periodicity $p_\text{A}$ or $p_\text{B}$, similar to the case in Fig.\ \ref{GRAll}(b). Slight shifts in the peak positions between the two traces are possibly due to the small difference in the electron density between different cooling downs. Some of the remnant features in Fig.\ \ref{GR100} are probably resulting from higher $n$ transverse resonances or from longitudinal resonances Eq.\ (\ref{BlAB}), although unambiguous identification turned out to be difficult as was the case for $a = 200$ nm.     

\section{Discussion}
\subsection{Towards artificial massless Dirac fermions\label{DiscMDF}}
Generation of artificial massless Dirac fermions (MDF) in conventional 2DEGs is an enticing possibility. Not only does it provide an alternative arena to pursue the physics of Dirac fermions, but it also offers an opportunity to precisely control or modify the properties of Dirac fermions; modern semiconductor nano-lithography technology, along with the length scale orders of magnitude larger than the inter-atomic distance, allows us to manipulate the hexagonal lattice at will. For instance, we will be able to introduce designed strain into the lattice to generate an effective magnetic field, \cite{GuineaNP10} or fabricate nano-ribbons truncated at the edges along desired orientations (zigzag or armchair), \cite{Nakada96} which will be extremely difficult to be performed on natural graphene.

Several attempts have been made to generate MDF in conventional GaAs/AlGaAs 2DEGs, \cite{Gibertini09,Simoni10,Nadvornik12} reporting (i) narrowing of a photoluminescence peak, \cite{Gibertini09} (ii) small amplitude magnetoresistance oscillations periodic in $B$ in the quantum Hall regime \cite{Simoni10} (qualitatively resembling those observed in the antidot lattices \cite{Iye04JPSJ}), and (iii) splitting of the cyclotron resonance absorption. \cite{Nadvornik12} The splitting in (iii) is similar to AM of SdHO in the present study in the sense that both directly reflect the broadening of the Landau levels due to the modulation (Landau bands). Although these phenomena undoubtedly stem from the potential modulation introduced into 2DEGs, they are not particularly sensitive to the lattice type of the 2D modulation. Above all, they are not probing the key ingredient in the generation of the artificial MDF, the formation of minibands by Bragg reflections. We believe that our observation of the open orbits engendered by Bragg reflections represents one step forward toward the realization of MDF\@.

However, a number of improvements are still needed to be made. (Criteria for achieving MDF in 2DEGs are discussed in detail in Ref.\ \onlinecite{Nadvornik12}). First, the electron density $n_e$ has to be reduced to $\sim n_\text{D} =(4/\sqrt{3}) a^{-2}$ in order to tune the Fermi energy $E_\text{F}$ close to the Dirac point. The density $n_\text{D}$ equals $6 \times 10^{13}$ and $2 \times 10^{14}$ m$^{-2}$ for $a = 200$ and 100 nm, respectively. For this purpose, we prepared HLSL devices equipped with a backgate which allows us to vary $n_e$. \cite{Kato12MT} However, the values of $n_\text{D}$ appear to be too small to be achieved with the 2DEG wafer used in the present study, preserving the quality (mobility) good enough for the minibands to be formed without being hampered by the disorder. It will be necessary to start with a 2DEG having a higher mobility and a much smaller $n_e$. Reducing the lattice constant $a$ will also be of help; by reducing $a$ to 50 nm, which is not impossible using our high-resolution resist, $n_\text{D}$ is augmented to a more amenable value of $9 \times 10^{14}$ m$^{-2}$. \cite{*[{Generally, a 2DEG wafer with a high mobility and a low electron density possesses a wide spacer layer that separates the 2DEG layer from the doping layer, and thus located at a large depth from the surface (typically $d \simeq 500$ nm, see, e.g., }][{), which is incompatible with the introduction of short length-scale modulation. One way to circumvent the problem is to resort to inverted structure that has the spacer and the doping layers beneath the 2DEG layer.}] Pfeiffer89} Second, in order to avoid the Dirac cones from energetically overlapping with other minibands, the sign of the hexagonal potential modulation ought to be positive (repulsive). \cite{Park09N,Nadvornik12,Gomes12} Unfortunately, none of the oscillations discussed in the present study (CO, AM of SdHO, and GROO) is sensitive to the sign of the modulation, and we therefore do not know whether our hexagonal potential modulation is repulsive or attractive. If the strain induced by the resist turns out to introduce attractive potential, it will be necessary to switch to the honeycomb lattice modulation dual to the hexagonal lattice. \cite{Gibertini09}  

Very recently, Gomes \textit{et al.} reported \cite{Gomes12} the generation of MDF on a 2DEG at a copper surface, using carbon monoxide molecules as a negative gate working on the 2DEG\@. The molecules were assembled by the atomic manipulation technique employing a scanning tunneling microscope (STM). They probed the resulting Dirac cones using the scanning tunneling spectroscopy. Although their result is impressive, their method requires acrobatic operation of the STM\@. Also, it will be probably not very easy to perform transport measurement on a 2DEG at the copper surface residing just above the bulk of the copper. We therefore believe that generating MDF in a conventional semiconductor 2DEG is a challenge still worth pursuing.

\subsection{Towards the observation of the bubble phase\label{DiscBP}}
We extended the magnetoresistance measurement in the dilution refrigerator (15 mK) up to 9 T, in search of the effect induced by the hexagonal modulation in the quantum Hall regime. We especially focused on 1/4 and 3/4 fillings of the $N = 2$ and higher Landau levels, seeking for signals related to the bubble phase theoretically predicted to be the ground state at these fillings. \cite{Koulakov96,Fogler96,Moessner96}

The bubble phase is a charge density wave (CDW) state in which clusters of two or three electrons are arranged in the hexagonal lattice. Experimentally, reentrant integer quantum Hall effect (RIQHE) observed at filling factors $\nu = 17/4$, 19/4,... was interpreted as resulting from the bubble state pinned by impurities. \cite{Cooper99} Later, microwave resonances were observed at these fillings, which were ascribed to the pinning mode of the bubble state. \cite{Lewis02} For these experiments, 2DEGs having a very high mobility ($\mu \agt 1000$ m$^2$V$^{-1}$s$^{-1}$) were required, since the formation of the fragile bubble state is readily quenched by disorder.
On the other hand, external modulation having the same lattice constant and the lattice type, namely the hexagonal lattice, as the bubble state is theoretically predicted to stabilize the bubble phase. \cite{Haldane00} We therefore expect the bubble states to be formed in optimally designed HLSLs even if the mobility is not as high. In fact, anisotropic magnetotransport was observed to be induced by external modulation in 1D lateral superlattices with a mobility $\mu \simeq 100$ m$^2$V$^{-1}$s$^{-1}$, which was ascribed to the stabilization of the stripe phase, \cite{Endo02f,Endo03lt} similarly fragile phase predicted to be the ground state at the half fillings. \cite{Koulakov96,Fogler96,Moessner96} 
Stabilization of the bubble phase by the external modulation in HLSLs would provide direct information on the spatial distribution of the charge through the lateral size and the shape of the introduced modulation. Note that experimental findings related to the bubble phase reported so far \cite{Cooper99,Lewis02} do not have sensitivity to the spatial distribution.  

Unfortunately, we have found no changes above 1 T attributable to the introduction of the modulation in HLSLs thus far for both $a = 200$ and 100 nm.
%Apparently, the quality of our 2DEG was not good enough to support the formation of the bubble phase, even with the aid of the hexagonal modulation. 
This is probably because either the introduced lattice constant $a$ or the strength of the potential modulation was not appropriate.
The lattice constant $a_\text{CDW}$ of the bubble state is theoretically predicted to be $a_\text{CDW} \sim 3 R_\text{c} = 3 \sqrt{2N+1} l$. \cite{Koulakov96,Fogler96}   For our $n_e = 3.9 \times 10^{15}$ m$^{-2}$, $a_\text{CDW} = 88$, 93, 98, and 102 nm for $\nu = 17/4$, 19/4, 21/4, and 23/4, respectively. Therefore, $a = 200$ nm appears to be rather too large to promote the formation of the bubble phase. On the other hand, $a = 100$ nm roughly matches the predicted $a_\text{CDW}$. In this case, however, the modulation was probably too weak to overcome the detrimental effect exerted by the disorder. 

Since the magnetic length $l$ hence the lattice constant $a_\text{CDW}$ at a fixed filling factor $\nu = n_e h / (eB)$ increases with decreasing $n_e$, we can increase the suitable $a$ hence the modulation amplitude by using smaller $n_e$.
Therefore, it will be desirable, here again, to prepare HLSL samples with 2DEGs with higher mobility and smaller $n_e$.

\section{Conclusions\label{Conc}}
We have observed three types of oscillatory phenomena in the magnetoresistance of hexagonal lateral superlattices (HLSLs): commensurability oscillations (CO), amplitude modulation (AM) of Shubnikov-de Haas oscillations (SdHO), and the geometric resonances of open orbits (GROO). Both CO and GROO contain components deriving from two periodicities, $p_\text{A} = \sqrt{3} a / 2$ and $p_\text{B} = a / 2$, immanent in the hexagonal lattice with the lattice constant $a$, while only the larger periodicity $p_\text{A}$ manifests itself in the AM of SdHO\@. As in the case of square or rectangular 2D lateral superlattices, amplitude of CO in HLSL is much smaller than that in 1D lateral superlattices having a similar strength of the potential modulation. By contrast, magnitude of AM in the SdHO is comparable to that in 1D lateral superlattices and represents the strength of the potential modulation correctly. The information obtained here on the landscape (characterized by the Fourier components and their amplitudes) of the hexagonal potential modulation will form the basis of understanding intriguing phenomena expected to be observed in the future studies. 

The observation of GROO reveals that minibands are generated by Bragg reflections from the hexagonal superlattices, which requires highly ordered modulation potential with a small enough lattice constant close to the Fermi wavelength. With further adjustment of the parameters of the 2DEG and of the modulation potential, it might become possible in the future to probe by magnetotransport experiments artificially designed massless Dirac fermions (MDF) generated in the miniband structure. 

\begin{acknowledgments}
% put your acknowledgments here.
This work was supported in part by Grant-in-Aid for Scientific Research (A) (18204029) and (C) (18540312) from the Ministry of Education, Culture, Sports, Science and Technology (MEXT).
\end{acknowledgments}

% Create the reference section using BibTeX:
\bibliography{ourpps,antidots,artgraphene,lsls,ninehlvs,magmod,twodeg,Graphene}

\end{document}